\documentclass[
reprint, 
superscriptaddress, 
frontmatterverbose,
nofootinbib, 
bibnotes, 
amsmath,amssymb, 
aps, pra, 
floatfix
]{revtex4-2}

\usepackage{dcolumn} 
\usepackage{bm} 
\usepackage{xr}
\usepackage{graphicx} 

\usepackage{xfrac}
\usepackage[nice]{nicefrac}
\usepackage{extdash}
\usepackage{epsfig}
\usepackage{isomath}
\usepackage{textcomp}
\usepackage{bm}
\usepackage[english]{babel}
\usepackage{todonotes}
\usepackage{color,soul}

\usepackage{times} 

\DeclareFontFamily{U}{euc}{}
\DeclareFontShape{U}{euc}{m}{n}{<-6>eurm5<6-8>eurm7<8->eurm10}{}%
\DeclareSymbolFont{AMSc}{U}{euc}{m}{n} 
\DeclareMathSymbol{\umu}{\mathord}{AMSc}{"16}
\newcommand{\ensuretext}[1]{\ensuremath{\text{#1}}}
\newcommand{\unit}[1]{\ensuretext{\textrm{\,}}\ensuremath{\mathrm{#1}}}

\newcommand{\eV}{\mathrm{eV}} 
\newcommand{\keV}{\mathrm{k}\eV} 

\newcommand{\Mum}{\ensuremath{\umu}\ensuremath{\mathrm{m}}} 
\newcommand{\mum}{\textrm{\,\ensuremath{\mathrm{\Mum}}}}

\mathchardef\ordinarycolon\mathcode`\:
\mathcode`\:=\string"8000
\begingroup \catcode`\:=\active
  \gdef:{\mathrel{\mathop\ordinarycolon}}
\endgroup

\begin{document}

\title{Heating in Multi-Layer Targets at ultra-high Intensity Laser Irradiation and the Impact of Density Oscillation}

\author{F. Paschke-Bruehl}
\affiliation{Helmholtz-Zentrum Dresden-Rossendorf, Bautzner Landstra{\ss}e 400, 01328 Dresden, Germany}

\author{M. Banjafar}
\affiliation{European XFEL, Holzkoppel 4, 22869 Schenefeld, Germany}

\author{M. Garten}
\affiliation{Helmholtz-Zentrum Dresden-Rossendorf, Bautzner Landstra{\ss}e 400, 01328 Dresden, Germany}
\affiliation{Technische Universit{\"a}t Dresden, 01069 Dresden, Germany}
\affiliation{now with Lawrence Berkeley National Laboratory, 1 Cyclotron Rd, Berkeley, CA 94720, United States}

\author{L.G. Huang}
\affiliation{Helmholtz-Zentrum Dresden-Rossendorf, Bautzner Landstra{\ss}e 400, 01328 Dresden, Germany}

\author{B.E. Marr\'e}
\affiliation{Helmholtz-Zentrum Dresden-Rossendorf, Bautzner Landstra{\ss}e 400, 01328 Dresden, Germany}

\author{M. Nakatsutsumi}
\affiliation{European XFEL, Holzkoppel 4, 22869 Schenefeld, Germany}

\author{L. Randolph}
\affiliation{European XFEL, Holzkoppel 4, 22869 Schenefeld, Germany}
\affiliation{Department Physik, Universit{\"a}t Siegen, 57072, Siegen, Germany}

\author{T.E. Cowan}
\author{U. Schramm}
\affiliation{Helmholtz-Zentrum Dresden-Rossendorf, Bautzner Landstraße 400, 01328 Dresden, Germany}
\affiliation{Technische Universit{\"a}t Dresden, 01069 Dresden, Germany}

\author{T. Kluge}
\affiliation{Helmholtz-Zentrum Dresden-Rossendorf, Bautzner Landstra{\ss}e 400, 01328 Dresden, Germany}

\date{\today}

\begin{abstract}

We present a computational study of isochoric heating in multi-layered targets at ultra-high intensity laser irradiation ($\sim 10^{20}\unit{W/cm}^2$). 
Previous studies have shown enhanced ion heating at interfaces, but at the cost of large temperature gradients. 
Here, we study multi-layered targets to spread this enhanced interface heating to the entirety of the target and find heating parameters at which the temperature distribution is more homogeneous than at a single interface while still exceeding the mean temperature of a non-layered target. 
Further, we identify a pressure oscillation that causes the layers to alternate between expanding and being compressed with non beneficial effect on the heating. 
Based on that, we derive an analytical model estimating the oscillation period to find target conditions that optimize heating and temperature homogeneity. 
This model can also be used to infer the plasma temperature from the oscillation period which can be measured e.g. by XFEL probing. 

\end{abstract}
\maketitle

\section{Introduction}
\vspace{-0.3cm}
Studying solids upon high-intensity laser irradiation allows to observe and learn about plasma dynamics, which have impact on a wide variety of applications, such as tumor therapy \cite{Kraft2010,Zeil2013,Karsch2017,Kroll2022}, fusion research \cite{Kodama2001,Roth2001,Fernandez2014,Zylstra2021} or laboratory astrophysics \cite{Bulanov2009,Bulanov2015a}. 
Previous studies have shown, that a specific target design can optimize the laser-target interaction towards preferred mechanisms like rear particle acceleration \cite{Snavely-IntenseProtonBeams,Maksimchuck2000}, front surface compression \cite{MacchiRPA,Mishra2009,Gaus2020} or bulk heating \cite{Chawla2013,Kluge2018a}. 
Following this approach, we present a computational study of multi-layered (ML) targets to study and enhance specifically the target heating. 
Our goal is to investigate the usage of ML targets to generate warm - or hot - dense matter in the embedded layers, a state of high, homogeneous temperature ( $> 1$ eV) and density ( $> 1$ kg/m$^3$).
Further, we observe complex dynamics of oscillating layer densities, occurring as a result of the layered target geometry and high temperatures. 
We present an analytic model for the oscillation period to predict optimal heating conditions, considering limiting quantities, such as ML layer thickness and density. \\
There are several processes known to transfer energy from the laser to bulk temperature, e.g. isochoric heating by particle acceleration at the front surface \cite{Patel2006,Roth2010,Higashi2020}, subsequent heat diffusion \cite{Higashi2020,Higashi2022}, shocks \cite{Silva2004,Akli2008}, or in the case of ultra-intense lasers also return current generation balancing the dense relativistic electron beam accelerated by the laser \cite{Glinsky1995a,Sherlock2014,Huang2016,Kluge2018a}. 
The latter dominates the heating for short time scales (100s fs) and has the advantage of relatively homogeneous heating through the target compared to diffusion or shocks. 
Nevertheless, heating by laser accelerated electrons has been found to be most efficient at interfaces, as derived in \cite{Huang2013}. 
The simulation study by Huang et al. (2013) has shown, that ion heating at a buried interface between two materials can increase the efficiency by more than a factor of two compared to a pure material, because of a TNSA-like (target normal sheath acceleration) expansion at the interface. 
However, the downside of such heating is an inhomogeneous heat distribution over the target, since only in the small volume - within a Debye length - around the interface does the enhanced heating occur. 
This motivates the present study to investigate ML targets, aiming at combining efficient \emph{and} homogeneous heating. 
In the ML target we stack multiple interfaces behind each other to spread the enhanced interface heating to the entire target depth.

\vspace{-0.5cm}
\section{Results}
\vspace{-0.2cm}
\subsection{Setup}
\begin{figure}
    \centering
    \includegraphics[width = 0.481\textwidth]{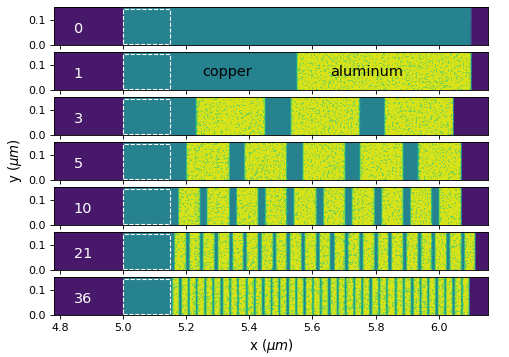}
    \vspace{-0.2cm}
    \caption{Target setup in the simulations with varying number of layers $n_l = \{0, 1, 3, 5, 10, 21, 36\}$ (top to bottom) and 150 nm copper shield (dashed, white line). Laser is incoming from the left.}
    \label{fig:Setup_CuAl_density}
\end{figure}
\begin{figure*}
    \centering
    \includegraphics[width = 0.88\textwidth]{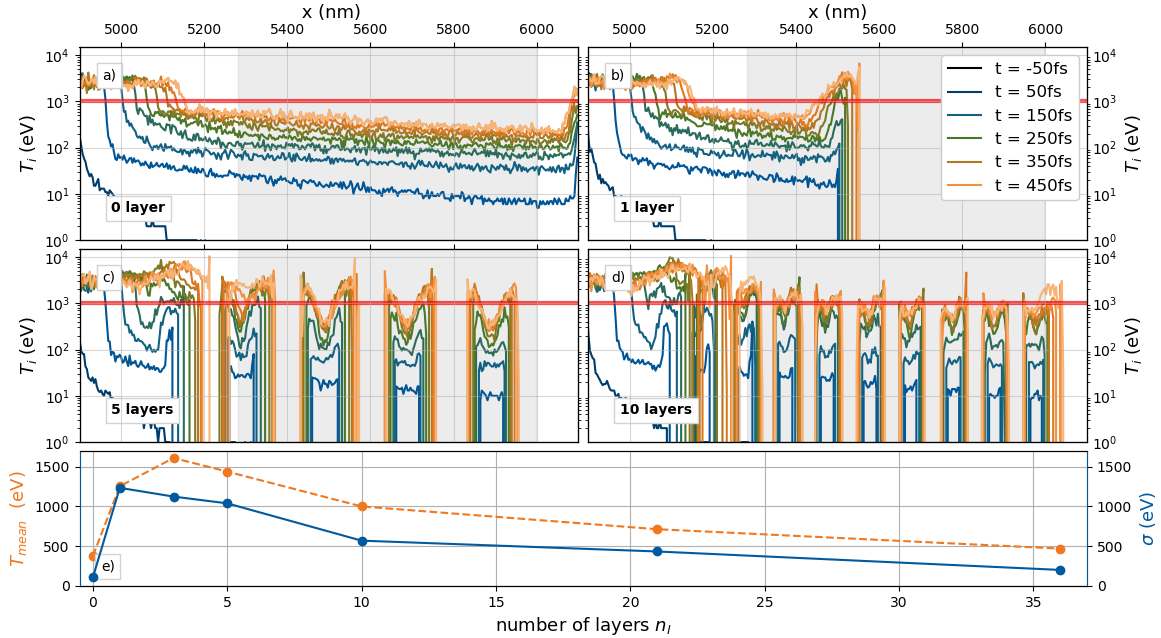}
    \caption{\textbf{a-d} Temperature of copper ions $T_i$ in the target over time for $n_l$ = \{0, 1, 5, 10\} layers with time relative to the arrival of the temporal peak of the laser pulse on the target surface. \textbf{e} Mean ion temperature (orange, dashed) and its variation (blue) in the ML region (shaded area). }
    \label{fig:temp_overview}
\end{figure*}
We performed 2D3V simulations (2 dimensional in space, 3 dimensional in velocity) to compare ML targets with different number of layers, as seen in Fig. ~\ref{fig:Setup_CuAl_density},  using the particle-in-cell code SMILEI \cite{Derouillat2018}. 
We focus on an exemplary target, made of copper and aluminum layers with a $150\unit{nm}$ thick front surface copper shield (cp. dashed white line in Fig. ~\ref{fig:Setup_CuAl_density}) to protect the ML structure and ensure similar laser absorption for all cases. 
While varying the number of layers $n_{l}$, the layer thicknesses are adjusted so that the total target thickness is kept roughly constant at $d_{tot} \simeq 1-1.1\unit{\mum}$.\\
The laser parameters in the simulations, if not stated otherwise, are peak intensity $I_L = 10^{20}\unit{W/cm}^{2}$, pulse duration $\tau_{FWHM} = 40 \unit{fs}$ (temporal gaussian profile, plane wave, linear polarization in y-direction), typical parameters e.g. for Ti:sapphire laser systems. 
Accordingly, we set the laser wavelength to $800\unit{nm}$.
See \emph{Methods} for more detailed computational parameters. 
All times will be given relative to the arrival of the laser pulse maximum on the target surface. \\
As we will discuss ion heating in the following sections, it is important to clarify the meaning of the term temperature $T_i$ in the simulation: as temperature we define the kinetic energy of ions derived from their momentum in z-direction only, $T_i = m_i p_z^2$. 
This dimension is not a spatial dimension, but a dimension in velocity only(2D3V simulation).
With the laser being polarized in y-direction, ions gain momentum in z-direction only via collisions. 
Hence, the temperature as defined here consists only of energy due to particle interactions and excludes directed energy caused by longitudinal or transverse movement, i.e. direct acceleration by the 2D laser. 
\begin{figure*}[ht]
    \centering
    \includegraphics[width = \textwidth ]{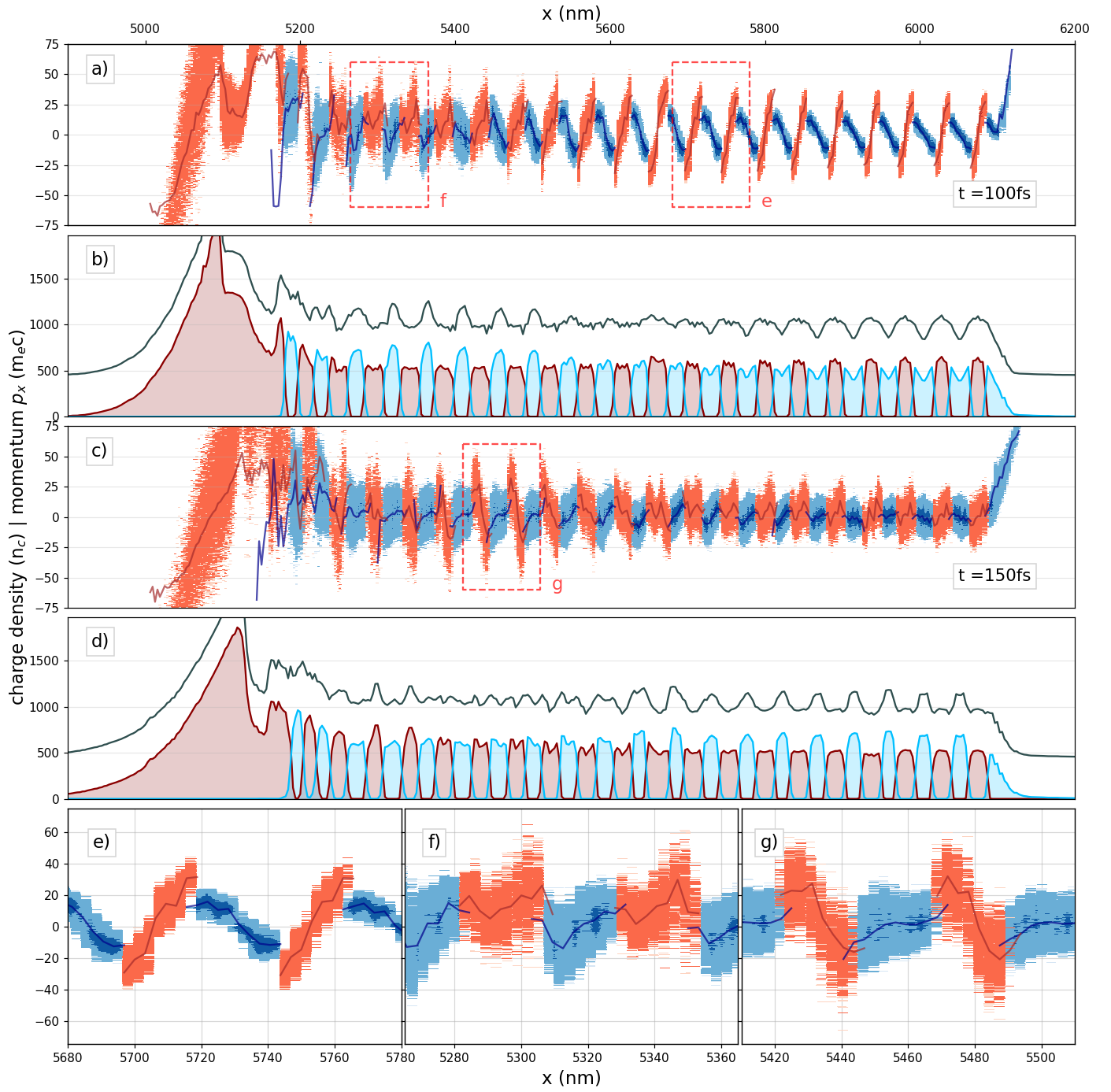}
    \caption{\textbf{a} Phase space of copper (orange) and aluminum (blue) at $t = 100\unit{fs}$ for $n_l =21$ with $p_x$ in laser direction. \textbf{b} Charge density at $t = 100\unit{fs}$ and free electron density (grey) with offset. \textbf{c} Phase space at $t = 150\unit{fs}$. \textbf{d} Charge density at $t = 150\unit{fs}$ and free electron density (grey) with offset. \textbf{e} Phase space at $t = 100\unit{fs}$. \textbf{f} Phase space at $t = 100\unit{fs}$. \textbf{g} Phase space at $t = 150\unit{fs}$. }
    \label{fig:phasespace}
\end{figure*}
\subsection{Heating in ML targets}
\label{heating}
First, we investigate the ion heating in the embedded copper layers in Fig.~\ref{fig:temp_overview}, showing the evolution of the copper temperature $T_i$ for a homogeneous target compared to ML targets with 1, 5 and 10 layers of each material, respectively. 
We observe an enhancement of temperature close to the interfaces for all cases, most dominant for the cases $n_l = 1$ and $n_l = 5$. 
This local interface heating occurs because of an expansion of the denser material (copper) into the less dense material (aluminum), induced by a pressure difference. 
The expansion mechanism is comparable to the well known process of TNSA, where electrons that expand into vacuum cause an electric field, which drag the ions with them. 
Here, ions expand into the lower density layer instead of vacuum, but also gain momentum, i.e. directed forward movement due to the electric field caused by the expanding electrons, as we see in detail in Fig.~\ref{fig:phasespace} and as predicted in \cite{Huang2013}. 
Due to the expansion and thus forward movement, ions collide, leading to a conversion of kinetic energy into thermodynamic heat and thus increased ion temperature compared to target regions where no expansion occurs and heating happens only by electron-ion collisions. 
For a larger number of layers the temperature starts to decrease, while the average throughout the target bulk always remains higher than in the homogeneous non-layered target (Fig.~\ref{fig:temp_overview}e, orange line). \\
In particular in Fig.~\ref{fig:temp_overview}e (blue line) we recognize that more layers cause a lower temperature spread through the target depth. 
This observation can be explained easily: more layers cause the expansion to happen all over the target and thus decrease the temperature spread between two regions, also seen by comparing Fig.~\ref{fig:temp_overview}b and d. \\
At the same time we observed a decrease in temperature when increasing the number of layers beyond $n_l=5$ for our specific target parameters. 
This heating deficit can be explained by a competing process of oscillating layer densities, which is the topic of the remainder of this paper. 
The oscillations cause the aforementioned expansions at the interfaces to decelerate and ultimately reverse due to reversing pressure relations of neighboring layers. 
This causes the particle momenta to reduce, and thus collisionality and heating to behave likewise. 
This leads to a decrease in temperature in comparison to a single interface, where no oscillation appears, i.e. where particles expand freely without reversion of momenta. \\
The optimum between the highest temperature (i.e. less interfaces) and a homogeneous temperature distribution (i.e. more interfaces) is around $n_l=10$ layers in our case, where the mean temperature is up by a factor of more than 2 compared to the homogeneous plasma and the temperature spread has already dropped to less than half of that in the single interface case.\\
In the following we will give an analytic description of the density oscillations in order to find a general expression for optimum parameters. 
\subsection{Density Oscillation}
In this section we want to understand and describe the aforementioned oscillation of layer densities in detail to further estimate its impact on the heating. Fig.~\ref{fig:phasespace} shows the phase space and density of copper (red) and aluminum (blue) over target depth for the case $n_{l} = 21$ and $t = 100\unit{fs}$ (a, b), as well as $t = 150\unit{fs}$ (c, d). 
In panels (e-g) we additionally zoom into the phase space behaviour to show the different stages of the copper layer, i.e. expansion (e), turning point (f) and compression (g). \\
The initial expansion of the copper layers, as in (e, orange), occurs because of the higher electron density and thus higher pressure compared to the neighboring aluminum layers. 
This simultaneously leads to a compression of the aluminum layers, see (e, blue), which causes them to increase in density. 
With the copper layer decreasing in density due to the expansion, the density and thus pressure relation between the materials slowly reverses, causing the dynamic to decelerate and invert, as seen in (f) and (g). 
The layers thus repeatedly oscillate in density - and also thickness - , the latter visible in panels (b) and (d). \\
As explained above, the reversion of the expansion, and thus ion momenta and collisionality, causes a heating deficit in comparison to a pure interface expansion. 
This leads to a condition for high plasma temperature: half of an oscillation period (i.e. the time to the first reversion of momenta) shall be large compared to the heating time, $T_{osc}/2 \gg \tau_{H}$, so that momenta do not reverse during the heating process and hence do not cause a heating deficit.
In the following we derive a simple analytic expression for the oscillation period $T_{osc}$ for a given material and layer thickness to later use in the heating context. 
First of all, the driving force of the oscillation is the pressure difference between layers, or specifically the difference in free electron density. The motion is caused by an overshooting, due to ion inertia, of the expanding/compressing motion compensating the pressure difference.\\
When looking at the density profiles in Fig.~\ref{fig:phasespace}, particularly the free electron density (grey) over time, one might think there is a global density motion through the target, comparable to a pressure wave, but that is not the case. 
As we will see in the following, the single layers are local oscillators, only dependent on local density and electron energy.    
\begin{figure}[ht]
    \centering
    \includegraphics[width = 0.47\textwidth]{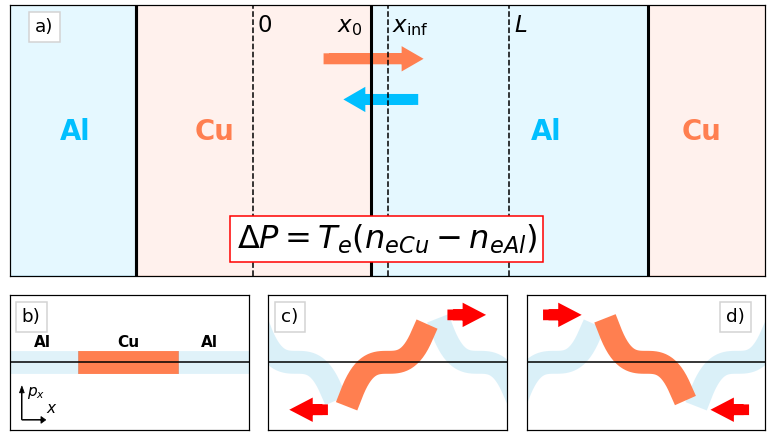}
    \caption{ 
        \textbf{a} Modeling of layers in the target and plasma pressure for Cu$|$Al target. We neglect the ion temperature when calculating the pressure difference. \textbf{b} Layer standing still. \textbf{c} Cu layer (orange) expands due to higher plasma pressure than neighboring aluminum layers (blue). \textbf{d} Cu layer is being compressed due to increased plasma pressure of neighbors. 
        \label{fig:model_scheme}}
\end{figure}
Since we are only interested in longitudinal forces ($x$-direction), we can reduce our analysis to 1 dimension. 
We start with the plasma pressure in two neighboring layers with index $j$ and $j+1$
\begin{equation} 
    \label{plasmaPressure1}
    \begin{aligned}
        P_{j}  &= P_{i,j} + P_{e,j} \\    
    \end{aligned}
\end{equation}
where 
\begin{equation}\\ \label{plasmaPressure}\\
    \begin{aligned}
        P_{i,j} = n_{i,j} T_{i} \\
        P_{e,j} = n_{e,j} T_{e} \\
    \end{aligned}
\end{equation}
are the ion and (free) electron pressure, respectively, with ion density $n_{i,j}$, free electron density $n_{e,j}\equiv n_j$ and respective temperatures $T_e$ and $T_i$ in layer $j$. \\
Initially, electrons are heated faster than ions ($T_{e}\gg T_{i}$), because of the mass relations ($m_i \gg m_e$). 
Therefore $P_{e,j} \gg P_{i,j}$ and we can neglect the pressure induced by ions. 
Eqn.~(\ref{plasmaPressure1}) then reduces to $P_{j} = P_{e,j}$. \\
Assuming the electron temperature to be similar in neighboring layers, because the temperature gradient in the target is large compared to the size of the layers, cp. ~Fig.(\ref{fig:temp_overview}), the pressure on the interface is then simply given by $\Delta P= \left(n_{j}-n_{j+1}\right)T_e$. 
This pressure leads to a compression of the low density layer (w.r.t. free electrons) and expansion of the high density layer. 
At some point the low density layer is compressed enough, that as the inert ions continue to move it reaches higher density than the other layer. 
This reverses the net pressure direction and the process starts again, leading to a continuous oscillation of the layers, see Fig.~\ref{fig:model_scheme}. \\
To describe this process quantitatively, we follow the derivation of the oscillation of two gases in a cylinder of length $L$ and separated by a heavy piston, as done by E.~Gislason \cite{Gislason2010}. 
We can adapt this model to our conditions by replacing the piston with the heavy bulk ions, so that the pressure primarily acts on them, while the pressure on the electrons is neglected due to their small mass. 
We now define the virtual ion piston, that can be thought of as the moving mass $M$, given by the mass of the \emph{moving} ions between the centers of two neighboring layers. 
The equation of motion for this virtual ion piston can then be written with the areal force density $\tilde f_x = \Delta P(t) = \Delta n(t) T_e$ with $\Delta n(t) = n_{j}(t) - n_{j+1}(t)$ as 
\begin{equation}
    \frac{d}{dt} \left( \tilde M X(t) \right) = \Delta n(t) T_e. 
    \label{eqn:eom1}
\end{equation}

Here, $\tilde M$ is the fraction $g_1 \tilde M_0$ of the total mass density $\tilde M_0 = x_0 \rho_j + \left(L-x_0\right) \rho_{j+1}$, $\ddot X(t) = g_2 \ddot x$ is the average acceleration at time $t$ of the ions, and $T_e$ is the electron temperature in x-direction. \\
As shown in Fig.~\ref{fig:model_scheme}, we can define the initial thickness of odd numbered layers $j$ is $2 x_0$ and that of even numbered layers $j+1$ is $2 \left( L-x_0 \right)$ and their interface at time $t$ is positioned at $x(t)$ with $x(0)=x_0$. 
The average free electron densities inside the respective layers are then connected with the momentary interface position by
\begin{equation} 
    \label{densitiesModel}
    \begin{aligned}
        n_j(t) &= n_{j}^0 \frac{x_0}{x(t)}\\
        n_{j+1}(t) &= n_{j+1}^0 \frac{L-x_0}{L-x(t)}.
    \end{aligned}
\end{equation}

\begin{figure}[ht]
    \centering
    \includegraphics[width = 0.482\textwidth]{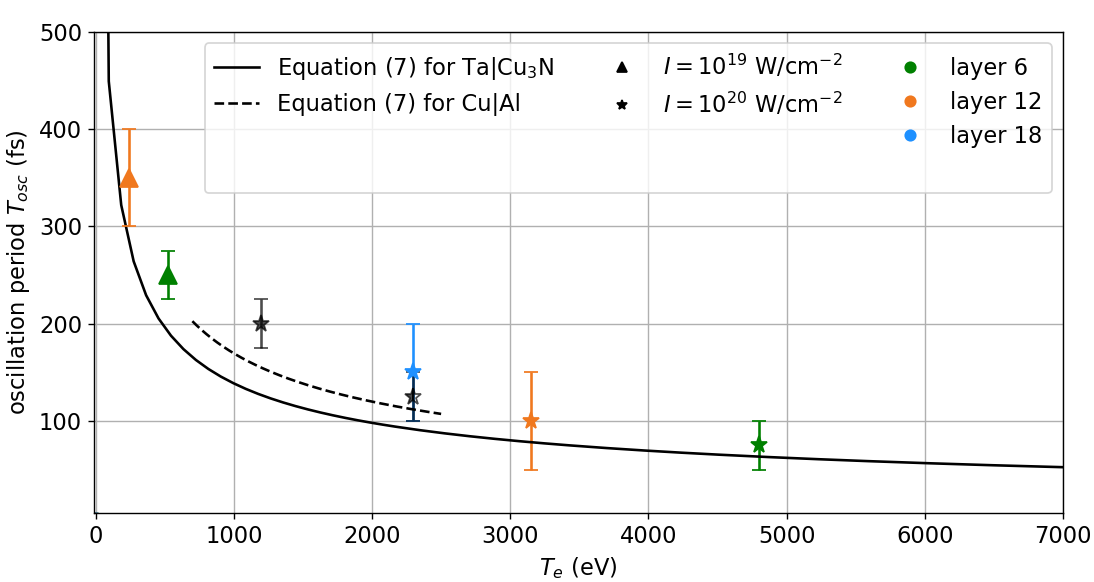}
    \caption{Comparison of oscillation period (Eqn.\ref{frequency}) and simulation results for tantalum and copper nitrite target. For the temperature $T_e$ only momenta in x-direction, i.e. momenta causing pressure in x-direction and thus at the interface, are considered. \label{fig:model_comparison}}
\end{figure}

To obtain $g_1$ and $g_2$, we have to recognize that possibly not all ions are moving ($ g_1$) and that not all ions that are moving do so with the same velocity ($g_2$). 
Regarding the former, for thin layer thicknesses as considered here, almost all ions are moving due to the comparatively large Debye length of the interface fields and we can set $g_1 \cong 1$. Only for thicker layers is $g_1 < 1$. 
Secondly, we have to take into account that those ions in motion are moving with a velocity depending on the distance from the center of the layer. 
The ions exhibit a velocity dispersion that is nearly linear, as can be seen in Fig.~\ref{fig:phasespace}.
Due to the symmetry of the problem, at the layer center the ion velocity is always zero, while at the interface it is at its maximum value $\dot x(t)$. 
We therefore apply the approximation for the average velocity of the moving ions at time $t$: $g_2\equiv \dot X/\dot x \cong 0.5$. \\
Eqn.~\eqref{eqn:eom1} can then be written as
\begin{equation} \label{equationofMotion}
    \tilde M_0 \ddot x \cong 2 T_{e} \left(n_j^0\frac{x_0}{x_\infty+\Delta x}-n_{j+1}^0\frac{L-x_0}{L-x_\infty-\Delta x}\right) 
\end{equation}
where $\Delta x\equiv x-x_\infty$ and $x_\infty$ is the equilibrium position where $P_j=P_{j+1}$, so $\tilde f_x = 0$ with Eqn.~\eqref{densitiesModel}, 
\begin{equation}
    \label{xinf}
    x_\infty = L \cdot \left[1 + \frac{n_{j+1}^0(L-x_0)}{n_{j}^0x_0}\right]^{-1}.
\end{equation}
We can now perform a Taylor expansion on the right hand side of Eqn.~\eqref{equationofMotion} for small values of $\Delta x\ll x_\infty$ and obtain  $\ddot x = -\Delta x \omega_{osc}^2$ with oscillation frequncy $\omega_{osc} = \frac{2\pi}{T_{osc}}$ and oscillation period $T_{osc}$. 
Then we obtain
\begin{equation} \label{frequency}
        \omega_{osc}^2 = \frac{2T_e}{\tilde{M}_0}\left[n_{j}^0\frac{x_0}{x_\infty^2} + n_{j+1}^0\frac{(L-x_0)}{(L - x_\infty)^2}\right].
\end{equation}
For thicker foils, where $g_1 < 1$, Eqn.~\eqref{frequency} resembles the lower limit for the oscillation frequency. \\
\begin{figure}[ht]
    \centering
    \includegraphics[width = 0.47\textwidth]{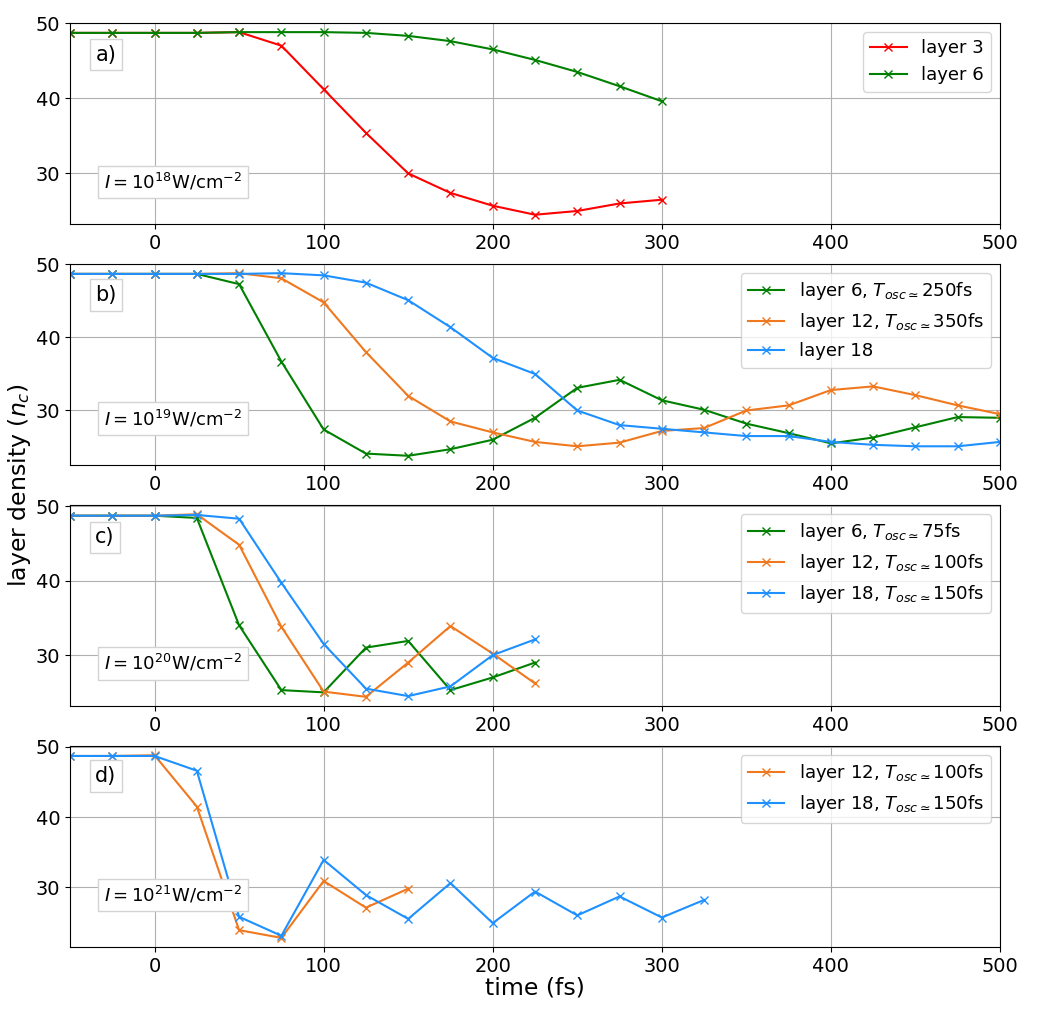}
    \caption{Density of tantalum layer over time for Ta$|$Cu$_3$N target and $n_l=21$ for different layer positions (green line at 6th layer (front), orange at 12th (center), blue at 18th (rear)) and laser intensities ((a) $10^{18}\unit{W/cm}^2$, (b) $10^{19}\unit{W/cm}^2$, (c) $10^{20}\unit{W/cm}^2$, (d) $10^{21}\unit{W/cm}^2$), both affecting the electron temperature in the layer. The density measurement is limited by the compression wave reaching the layer. 
    \label{fig:OscillationPeriods}}
\end{figure}
The oscillation frequency as derived above depends on three fundamental quantities: the layer thicknesses, the layer composition (ion/electron densities and masses) and the electron temperature.\\
With Eqn.~\eqref{frequency} the oscillation period is expected to increase with increasing areal mass density and decrease with $T_{e}$, so the layers oscillate slower for heavier materials, but faster for higher temperatures. To further investigate the latter dependency we performed additional simulations, this time varying laser intensity. 
These simulations contain tantalum and copper nitrite layers, similar geometry to the setup in Fig.~\ref{fig:Setup_CuAl_density} for $n_l = 21$.\\
Fig.~\ref{fig:OscillationPeriods} shows the density over time for different tantalum layer, varying either their position in the target or the laser intensity and thus the local electron temperature. We see that higher laser intensities and positions further to the front, i.e. higher $T_e$, cause shorter oscillation periods, as expected from Eqn.~\eqref{frequency}. Further, we can measure $T_{osc}$ and compare to Eqn.~\eqref{frequency}, as shown in Fig.~\ref{fig:model_comparison}. 
There we show the oscillation period in dependence of the temperature. 
The agreement of the analytic model and the simulations is surprisingly good, considering that we take only the average layer motion into account and neglect any damping terms in Eqn.~(\ref{frequency}). 

\subsection{Optimal Heating}

We can now use the derivations above to obtain an approximate condition for optimal heating in a multi-layered target. 
First, we require that the expansion of the layer of interest shall be equal or larger than half the neighboring layer thickness, i.e. 
\begin{equation}
    c_s \tau > x_0 
    \label{eqn:condition1}
\end{equation}
so that the enhanced heating happens in the entirety of the layer and thus ensure homogeneous heating. 
Here, $c_{s}=\left(T_i Z/M_j\right)^{1/2}$ is the ion sound speed and $\tau$ is the expansion time, i.e. the smaller of either half of the oscillation period or (for thick layers) the heating time, $\tau = \min\left[T_{osc}/2,\tau_H\right]$.\\
In addition to that we require half of the oscillation period, i.e. until the first turning point, to exceed the heating time to ensure an advantage in heating in comparison to a non-layered target, as discussed before. This brings 
\begin{equation}
    T_{osc}/2 > \tau_{H}. 
    \label{eqn:condition2}
\end{equation}
We can now make an estimation for the copper and aluminum target above and compare to Fig.~\ref{fig:temp_overview}. 
The average bulk Cu temperature for our laser conditions is approx. $T_i = 1.1\unit{\keV}$ and with $Z/M_{Cu}\approx (2 m_p)^{-1}$ condition \eqref{eqn:condition1} can be evaluated to $d_{Cu} \le 46\unit{nm}$. 
We defined $d_{Al} = 2.6 d_{Cu}$, hence the only free parameter is $L$, or in other words $n_l$.
In the $850\unit{nm}$ thick ML region of the target we therefore require more than $n_l=5$ layers of each copper and aluminum. 
At the same time condition \eqref{eqn:condition2} can be evaluated for an approximate heating time in our conditions of $\tau_H\approx 200\unit{fs}$ (cp. Fig.~\ref{fig:temp_overview}) and Eqn.~\eqref{frequency} to $n_l \le 8$. 
As can be seen in Fig.~\ref{fig:temp_overview}e, this is exactly the range where the homogeneity of the temperature rapidly improves while the average temperature is still much higher than in the homogeneous target. 

\section{Discussion and Conclusion}
We performed simulations of ML targets and studied the isochoric heating in the embedded layers. Further, we successfully expanded the enhanced interface heating, as stated in \cite{Huang2013}, to the entirety of the target and discovered a limiting effect on the heating - the density oscillation. The derived analytic expression for the oscillation period shows good agreement with the 2D3V particle-in-cell simulations. 
Based on the analytic expression for the oscillation period, we can derive a ML target setup for optimal heating, exceeding the temperatures of a homogeneous target, containing only one material.\\
Finally, we want to highlight that the ML structure is not only advantageous for enhanced heating of the embedded layers, but it can also serve as a diagnostic tool e.g. for GISAXS (grazing-incidence small-angle x-ray scattering), as done in \cite{Randolph2022}. 
GISAXS allows to estimate the densities in the layers, as the x-ray scattering pattern changes with the layer quantities, so one could infer the oscillation frequency and thus calculate the local plasma temperature without any further assumptions.
Though the temperature can be inferred spectroscopically or from X-Ray Thomson Scattering (XRTS) including novel, essentially model free schemes for homogeneous plasmas \cite{Dornheim2022}, those methods rely on complicated atomic physics in modeling the radiative properties of warm and hot dense matter (spectral methods), or average over the full depth of the target (XRTS). 
Hence a submicron temperature resolution needs careful future theoretic modelling and is beyond the scope of the current state-of-the-art.
Here, buried ML structures as proposed above could pose a possible alternative. 
Buried ML structures at different depths could be used to investigate the depth heating profile with GISAXS. 

\section*{Acknowledgements}

\section*{Data Availability}
The simulation input files and produced raw data is available under https://rodare.hzdr.de/ 

\section*{Author Contributions} Conceptualization and writing - original draft preparation F.P.-B., T.K.; draft review T.K., B.E.M., M.G.; simulations and data-analysis F.P.-B.; subject specific comments and support L.R, L.G.H, M.N., M.B.; project administration T.E.C., U.S. All authors have read and agreed to the published version of the manuscript. 

\section*{Methods}
\textbf{Computational Parameters}
\label{parameters}
The spacial simulation geometry is semi 2D, describing a two dimensional (x,y) box with very small spread in transverse direction. The simulation box measures 0.15x10 $\mu$, i.e. 48 cells in x direction (laser-longitudinal), 3200 cells in y direction (laser-transverse) and a resolution of 256 cells per laser wave length ($\lambda = 0.8 \mu$m). According to the CFL-condition (Courant–Friedrichs–Lewy), the time resolution was set to 356 time steps per laser period. Further, we assigned 65 particles per cell for each material (electrons, ions).
\renewcommand\bibname{Bibliography}


%

\end{document}